# Interpretation of interconnection of alternative macroscopic approaches associated with the Abraham-Minkowski controversy using microscopic description

O. B. Vorobyev

*Abstract* – **interconnection of alternative equations of conservation of the momentum related with different interpretations of parameters of equations of the macroscopic electromagnetic field is an essential aspect of the Abraham-Minkowski controversy, which is associated in the first place with rivalry of alternative definitions of optical momentum of the electromagnetic wave in a medium. Assuming physical significance of alternative equations of momentum conservation in accord with their interconnection, potential functional equivalence of alternative descriptions of optical momentum is examined in the context of the radiation pressure. In accordance with conducted examination, the material derivative of the momentum of a closed medium-field system is preserved in the case of transformations linking alternative equations of momentum conservation, whereas the time derivative of optical momentum as well as optical momentum is changed in compliance with alternative approaches. It is found that Abraham's, Minkowski's, and the total momenta are inherently linked with specific ponderomotive forces, which are not interchangeable in spite of interconnection of involved forms of field equations and equations of conservation of the momentum. Interconnection of alternative macroscopic approaches is clarified using microscopic interpretation of the energy and the momentum of the electromagnetic wave in a host medium. Contradictory references to microscopic aspects associated with alternative approaches, including assumptions in regard to manifestation of the Lorentz force on the level of macroscopic description and nature of inertial properties of electromagnetic wave in a host medium, are evaluated in this context. It leads to clarification of properties of ponderomotive forces as well as concepts of momentum and pseudomomentum of the electromagnetic wave in a host medium, which are essential for explanation of interconnection of alternative macroscopic approaches.**

## 1. Introduction

Description of mechanical properties of the electromagnetic wave in a host medium is a significant problem [1] - [19]. It is known as the Abraham-Minkowski controversy associated in the first place with the Abraham $\int \boldsymbol{E} \times \boldsymbol{H} dV$ and Minkowski $\int \boldsymbol{D} \times \boldsymbol{B} dV$ momenta, which have been interpreted as alternative electromagnetic momenta of the electromagnetic wave in a host medium [1] – [3]. Both of them are backed up by interpretations of empirical evidence [5] - [10] as well as equations of conservation of the momentum based on equations of the macroscopic electromagnetic field.

Besides, a number of alternative equations of conservation of the momentum related with assumptions in regard to separation of medium and electromagnetic parameters [8] - [21] of equations of the macroscopic electromagnetic field have been derived as well. In accord with those assumptions, the Abraham momentum of the electromagnetic field is transformed into a mechanical component of op-tical momentum [7] by means of the Lorentz force associated with microscopic description of medium-field interaction. In this case, optical momentum of the electromagnetic wave in a host medium has been interpreted as a combination of electromagnetic and "mechanical" momenta known as the total momentum of the electromagnetic wave in a host medium [18] - [19]. At the same time, the Abraham force singled out from the time derivative of the Minkowski momentum, as in the case of the Abraham approach, has been also interpreted as the time derivative of a "mechanical" component of optical momentum [5], [9]. Thus, nature of optical momentum of the electromagnetic wave in a host medium [22] as well as its form are aspects of the Abraham-Minkowski controversy.

Because alternative equations of conservation of the momentum are derived using equations of the macroscopic electromagnetic field [10], those equations are inherently interconnected. Interconnection of the equations of conservation of the momentum of the macroscopic electromagnetic field has been explained by absence of unique prescription for distribution of electromagnetic and medium parameters of momentum conservation equations [8], [9]. Such interpretation of interconnection assumes that alternative approaches represent correct descriptions [8], application of which is justified by their convenience in boundaries of specific physical problems [5].

Using an assumption that interaction of the electromagnetic field with atoms of a homogeneous medium provides zero contribution to optical momentum, it has been suggested that optical momentum represents an electromagnetic momentum, which takes forms of the Abraham and Minkowski momenta [16] in different conditions linked with radiation pressure. Another interpretation of the radiation pressure [11] has been recently advocated in the context of discussion of interconnection of alternative approaches. In this case, the average radiation pressure on an interface between dielectrics is explained in the framework of rival optical momenta by action of the Helmholtz force because contribution of the Abraham force is not significant due to fast oscillations of optical fields.

Nevertheless, in accordance with discussion of the alternative macroscopic approaches in the Section 2, interconnection of the alternative optical momenta of the electromagnetic wave is not reduced to redistribution of medium and electromagnetic components of equations of conservation in general case. It is demonstrated that while the material derivative of the momentum of a closed medium-field system is preserved in the case of transformations connecting alternative equations of conservation, the time derivative of optical momentum as well as optical momentum is changed in accordance with alternative approaches. At the same time, because discussed alternative equations are related with different forms of equations of the macroscopic electromagnetic field, it is difficult to justify ranks of significance of alternative momenta and corresponding ponderomotive forces on the level of macroscopic description.

Potential functional equivalence of alternative concepts of the optical momentum is examined in the case of description of the radiation pressure on an interface between dielectrics in the Section 3. It is demonstrated that averaging of oscillations of the Abraham force at optical frequencies does not eliminate differences of alternative approaches. It is obtained that the time-averaged radiation pressure is determined by dependence of optical momentum on the refractive index, which is defined by properties of medium components of alternative optical momenta. However, causal mechanism of conservation of the material derivative of a closed medium-field system is not clear in the context of alternative definitions of optical momentum of the electromagnetic wave in a host medium.

The alternative optical momenta associated with different presentations of field equations [4] - [10], [16] - [17] have been linked with contradictory interpretations of medium properties. Properties of components of optical momentum depending on the polarization and magnetization of a host medium, which are in the center of interconnection of alternative approaches, are discussed in the Section 4 using microscopic description of electromagnetic wave interaction with a host medium. Making use of the microscopic description, which takes into account secondary radiation of bound electrons of atoms of a host medium, it is demonstrated that the electromagnetic wave in a host medium is accompanied by "mechanical" and "electromagnetic" components of the pseudomomentum [22], which are intrinsically interconnected. In contrast with properties of the electromagnetic wave in free space as well as matter particles, momenta of which are determined by the velocity of translational motion of corresponding relativistic masses, components of optical pseudomomentum related with polarization and magnetization of a host medium depend on interaction of the electromagnetic wave with a host medium, which properties are essential for description of inertial properties of the electromagnetic wave in a host medium. The material derivative of "electromagnetic" and "mechanical" components of optical momentum associated with medium-field interaction is interpreted as a generalized Lorentz force, which is responsible for momentum exchange between the electromagnetic wave and a host medium. Thus, interconnection of alternative approaches, Abraham's and Minkowski's forms of momenta as well as causal mechanism of conservation of the material derivative of a closed medium-field system are explained using microscopic description of interaction of the electromagnetic wave with a host medium.

## 2. ALTERNATIVE APPROACHES AND THEIR INTERCONNECTION

### 2.1. THE MINKOWSKI MOMENTUM

The macroscopic electromagnetic field is described by following equations [1], [2], [23]

$$\nabla \cdot \boldsymbol{D} = \rho, \qquad \nabla \times \boldsymbol{E} = -\frac{\partial \boldsymbol{B}}{\partial t}, \qquad (1)$$
$$\nabla \cdot \boldsymbol{B} = 0, \qquad \nabla \times \boldsymbol{H} = \frac{\partial \boldsymbol{D}}{\partial t} + \boldsymbol{j},$$

where $\rho$ and $\boldsymbol{j}$ are densities of free charges and electrical current, respectively; $\boldsymbol{D}$ is the displacement of the electric field $\boldsymbol{E}$, and $\boldsymbol{B}$ is the induction of the magnetic field $\boldsymbol{H}$. The field and auxiliary vectors are connected by constitutive relations $\boldsymbol{D} = \varepsilon_0 \varepsilon \boldsymbol{E}$, $\boldsymbol{B} = \mu_0 \mu \boldsymbol{H}$, where $\varepsilon_0$ and $\mu_0$ are the permittivity and the permeability of free space, $\varepsilon$ and $\mu$ are the relative permittivity and the permeability of an isotropic medium, respectively. It is worth noting that while description of medium properties based on constitutive relations is totally adequate for calculation of fields, it is not sufficient for unique description of the energy and the momentum of the electromagnetic field [24] - [26] in accord with alternative interpretations of equations of the macroscopic electromagnetic field reviewed e.g. in [10].

The momentum continuity equation for Minkowski's momentum can be deduced [5] using summing of equations, which are obtained by multiplication of scalar equations in the first column (1) by fields $\boldsymbol{E}$ and $\boldsymbol{H}$ and vector multiplication of equations in the second column by inductions $\boldsymbol{D}$ and $\boldsymbol{B}$ respectively,

$$\frac{\partial}{\partial t} \boldsymbol{D} \times \boldsymbol{B} + \nabla \cdot \bar{\bar{T}}^{Min} = -\boldsymbol{f}^{Lf}, \qquad (2)$$

where $\boldsymbol{D} \times \boldsymbol{B}$ is the density of Minkowski's momentum, significance of which for description of the momentum transferred by the electromagnetic wave to a body in a host medium has been justified by a number of experiments [8], $\boldsymbol{f}^{Lf} = \rho \boldsymbol{E} + \boldsymbol{j} \times \boldsymbol{B}$ is the density of the Lorentz force responsible for dissipation of the energy and momentum of the electromagnetic field by free charges [23], and $\bar{\bar{T}}^{Min}$ is the Minkowski stress tensor. The divergence of the stress tensor $\bar{\bar{T}}^{Min}$ is determined by this formula

$$\nabla \cdot \bar{\bar{T}}^{Min} = \boldsymbol{D} \times (\nabla \times \boldsymbol{E}) + \boldsymbol{B} \times (\nabla \times \boldsymbol{H})$$
$$-\boldsymbol{E}(\nabla \cdot \boldsymbol{D}) - \boldsymbol{H}(\nabla \cdot \boldsymbol{B}). \quad (3)$$

Using relations (1) - (3), the Minkowski stress tensor is presented in the form [5]

$$T_{ij}^{Min} = -(E_i D_j + H_i B_j) + \frac{1}{2}\delta_{ij}(\boldsymbol{E} \cdot \boldsymbol{D} + \boldsymbol{H} \cdot \boldsymbol{B}), \quad (4)$$

where $U = \frac{1}{2}(\boldsymbol{E} \cdot \boldsymbol{D} + \boldsymbol{H} \cdot \boldsymbol{B})$ is the total energy density of the electromagnetic wave in a host medium with negligibly small dispersion and losses in accordance with Abraham's and Minkowski's approaches.

Because the density of Minkowski's momentum depends on the refractive index of a host medium $n = \sqrt{\varepsilon\mu}$, momentum conservation equation (2) has to be complemented by the Helmholtz force density [23] $\boldsymbol{f}^H = -\frac{1}{2}\{\varepsilon_0 E^2 \nabla\varepsilon + \mu_0 H^2 \nabla\mu\}$ in the case of an inhomogeneous medium

$$\frac{\partial}{\partial t}\boldsymbol{D} \times \boldsymbol{B} + \nabla \cdot \bar{\bar{T}}^{Min} = -\boldsymbol{f}^{Lf} - \boldsymbol{f}^H. \quad (5)$$

The Helmholtz force density describes the rate of change of the density of medium momentum associated with medium inhomogeneity while it is not a component of optical momentum of the electromagnetic wave in a host medium.

Using that the phase $\boldsymbol{v} = \frac{c\hat{\boldsymbol{s}}}{n}$, group, and energy velocities of the electromagnetic wave [23] - [25] coincide in the case of a non-dispersive medium with negligible losses, the energy density of the electromagnetic wave $U$ is determined by this formula

$$U = \frac{n}{c}S, \quad (6)$$

where $\boldsymbol{S} = \boldsymbol{E} \times \boldsymbol{H}$ is the energy current density, $S = \hat{\boldsymbol{s}} \cdot \boldsymbol{S}$, $\hat{\boldsymbol{s}}$ is the unit vector in the direction of the energy current, $\varepsilon_0 \mu_0 = c^{-2}$, and $c$ is the velocity of light in free space. It is worth noting that $U$ is the energy density of the electromagnetic field associated with the electromagnetic momentum in accordance with Minkowski's approach.

The stress tensor is presented in the form $T_{ij}^{Min} = \frac{n^2}{c^2}S_i v_j$ [27], [28] using the energy current density of the electromagnetic wave and the density of Minkowski's momentum, $\frac{n^2}{c^2}\boldsymbol{S}$. While the divergence of the stress tensor, $\nabla \cdot \bar{\bar{T}}^{Min}$, corresponds to the gradient of the magnitude of the flux density of Minkowski's momentum, $\nabla \frac{n}{c}S$, in the case of the electromagnetic wave in a non-dispersive medium with negligible losses, the equation of conservation (5) is reduced to an equation $\frac{\partial}{\partial t}\boldsymbol{D} \times \boldsymbol{B} + \nabla \frac{n}{c}S = -\boldsymbol{f}^H$. The left-hand side of this equation represents the Lagrangian derivative of density of Minkowski's momentum while the force density $\boldsymbol{f}^H$ represents the material derivative of the mechanical momentum of a host medium, which motion is non-essential due to a negligibly small velocity of mechanical waves. It demonstrates that the latter equation as well as equation (5) describes conservation of the density of the full momentum of a closed medium-field system, which represents the sum of electromagnetic wave's momentum and the mechanical momentum of a host medium.

## 2.2. THE ABRAHAM MOMENTUM

In the case of the Abraham approach [2], [27], the time derivative $\frac{\partial}{\partial t} \boldsymbol{D} \times \boldsymbol{B}$ of Minkowski's form of the density of optical electromagnetic momentum is interpreted as the sum of the time derivative of the density of Abraham's electromagnetic momentum depending on field vectors and the density of the Abraham force related with medium parameters. Using the density of the Abraham momentum, $\varepsilon_0 \mu_0 \boldsymbol{E} \times \boldsymbol{H}$, the Abraham force density is presented through the use of the medium polarization, $\boldsymbol{P}$, and the medium magnetization, $\boldsymbol{M}$,

$$\boldsymbol{f}^A = \frac{\partial}{\partial t}\{\boldsymbol{D} \times \boldsymbol{B} - \varepsilon_0 \mu_0 \boldsymbol{E} \times \boldsymbol{H}\} = \frac{\partial}{\partial t}\left\{\boldsymbol{P} \times \boldsymbol{B} - \frac{1}{c^2}\boldsymbol{M} \times \boldsymbol{E}\right\}, \tag{7}$$

where it is used that $\boldsymbol{D} = \boldsymbol{P} + \varepsilon_0 \boldsymbol{E}$, and $\frac{\boldsymbol{B}}{\mu_0} = \boldsymbol{M} + \boldsymbol{H}$.

Making use of the Abraham force density, one can present the Abraham equation of conservation of the momentum in a following form

$$\frac{1}{c^2}\frac{\partial}{\partial t}\boldsymbol{S} + \nabla \cdot \bar{\bar{T}}^{Abr} = -\boldsymbol{f}^{Lf} - \boldsymbol{f}^H - \boldsymbol{f}^A, \tag{8}$$

where $\bar{\bar{T}}^{Abr}$ is Abraham's stress tensor, whose elements are determined in the frame of reference connected with a host medium by the symmetrized expression $\frac{1}{2}\left(T_{ij}^{Min} + T_{ji}^{Min}\right)$ [27].

In the case of a homogeneous medium, Abraham's and Minkowski's stress tensors are identical. Nevertheless, the density of the Abraham momentum $\frac{1}{c^2}\boldsymbol{S}$ corresponds to Abraham's photon momentum, which is determined by the expression $\frac{\hbar\omega}{cn}$ in the case a non-dispersive medium, whereas the density of the Minkowski momentum, $\frac{n^2}{c^2}\boldsymbol{S}$, corresponds to a different form of the photon momentum, $\frac{n\hbar\omega}{c}$, in a host medium. Comparison of those forms of optical momenta in the context of the equation $T_{ij}^{Min} = \frac{n^2}{c^2}S_i v_j$ discussed in the previous subsection demonstrates discrepancy between Abraham's momentum and its flux [6] in the case of the equation (8).

While the Abraham momentum is interpreted as the optical momentum of the electromagnetic wave [16] in a host medium similar to the case of the Minkowski momentum, the equation of conservation of the momentum (8) takes form

$$\frac{1}{c^2}\frac{\partial}{\partial t}\boldsymbol{S} + \nabla \cdot \left(\bar{\bar{T}}^{Abr} - \bar{\bar{T}}^A\right) = -\boldsymbol{f}^{Lf} - \boldsymbol{f}^{Af}, \tag{9}$$

where $T_{ij}^{Abr} - T_{ij}^A = \frac{1}{c^2}S_i v_j$ is a stress tensor corresponding to the flux density of the Abraham momentum of the electromagnetic wave, $\bar{\bar{T}}^A$ is a stress tensor corresponding to the flux density of the momentum, the time derivative of which is determined by the Abraham force, and $\boldsymbol{f}^{Af} = \nabla \cdot \bar{\bar{T}}^A + \boldsymbol{f}^H + \boldsymbol{f}^A$ is a force density related with the radiation pressure associated with Abraham's momentum.

In the case of another interpretation, the Abraham force is the time derivative of a mechanical momentum [8], [9], which accompanies Abraham's momentum. In this case, equations (8) and (9) are transformed into an alternative equation

$$\frac{1}{c^2}\frac{\partial}{\partial t}\boldsymbol{S} + \boldsymbol{f}^A + \nabla \cdot \bar{\bar{T}}^{Abr} = -\boldsymbol{f}^{Lf} - \boldsymbol{f}^H, \qquad (10)$$

where $\frac{1}{c^2}\boldsymbol{S} + \int \boldsymbol{f}^A dt$ is the density of a compound momentum of the electromagnetic wave, which has the form of Minkowski's momentum. It is worth noting that the mechanical component of the optical momentum $\int \boldsymbol{f}^A dt$ is not equal to zero, even though the time averaged Abraham force is vanishing due to oscillations of the function $\int \boldsymbol{f}^A dt$.

The difference of equations (10) and (9) is presented by a following equation

$$\boldsymbol{f}^A + \nabla \cdot \bar{\bar{T}}^A = -(\boldsymbol{f}^H - \boldsymbol{f}^{Af}), \qquad (11)$$

It is worth noting that the expression $\boldsymbol{f}^A + \nabla \cdot \bar{\bar{T}}^A$ represents the Lagrangian derivative of the medium related momentum of the electromagnetic wave because the Abraham force is the time derivative the momentum of the electromagnetic wave related with the medium polarization and magnetization. Correspondingly, the force density $\boldsymbol{f}^H - \boldsymbol{f}^{Af}$ takes into account medium inhomogeneity in this context while $\boldsymbol{f}^A = -\nabla \cdot \bar{\bar{T}}^A$ in the case of a homogeneous medium. In contrast, the medium related component of optical momentum is set to be zero in the case of the equation (9) while terms of the equation (11) are jointly eliminated.

Therefore, using equivalent transformations of corresponding continuity equations, the Abraham force can be interpreted as the time derivative of the mechanical momentum, which accompanies the electromagnetic momentum in accordance with the equation (10) or the time derivative of the part of Minkowski's electromagnetic momentum in accordance with the equation (2). The combination of forces linked by the equation (11) is eliminated in the case of equation (9), when the Abraham momentum is interpreted as the optical momentum of the electromagnetic wave in a host medium.

### 2.3. THE TOTAL MOMENTUM

The total momentum refers in the paper to the concept of electromagnetic wave momentum, whereas the full momentum [22] of a closed medium-field system is comprised of an optical momentum of the electromagnetic wave in a host medium and a mechanical momentum of a host medium linked with radiation pressure. The concept of the total momentum of the electromagnetic wave in a host medium assumes a particular interpretation of medium related momentum accompanying the electromagnetic wave in a host medium. In contrast with the equation (9), only a part of the momentum, the time derivative of which is determined by the Abraham force, is eliminated in the case of the total momentum, whereas its rest is interpreted as the "mechanical" momentum accompanying the electromagnetic wave [18] - [19]. The time derivative of the "mechanical" momentum of the electromagnetic wave in a host medium is determined in this case by Lorentz's type force.

Variants of the total momentum concept are linked with Einstein-Laub's, Chu's, and Ampere's equations of conservation of the momentum [10], which are similar in the case of a dielectric medium. Einstein-Laub's momentum conservation equation is presented in the form [3], [10]

$$\frac{1}{c^2}\frac{\partial}{\partial t}\boldsymbol{S} + \nabla \cdot \bar{\bar{T}}^{EL} = -\boldsymbol{f}^{EL}, \qquad (12)$$

where $T_{ij}^{EL} = -(E_i D_j + H_i B_j) + \frac{1}{2}\delta_{ij}(\varepsilon_0 E^2 + \mu_0 H^2)$ is the Einstein-Laub stress tensor and $\boldsymbol{f}^{EL}$ is the density of the E-L force, which is interpreted as Lorentz's type force. The force density $\boldsymbol{f}^{EL}$ is presented by the following formula

$$\boldsymbol{f}^{EL} = \boldsymbol{f}^{LfLE} + (\boldsymbol{P}\cdot\nabla)\boldsymbol{E} + \mu_0(\boldsymbol{M}\cdot\nabla)\boldsymbol{H} + \mu_0\frac{\partial \boldsymbol{P}}{\partial t}\times\boldsymbol{H} + \frac{1}{c^2}\boldsymbol{E}\times\frac{\partial \boldsymbol{M}}{\partial t}, \tag{13}$$

where $\boldsymbol{f}^{LfLE} = \rho\boldsymbol{E} + \boldsymbol{j}\times\mu_0\boldsymbol{H}$, $\boldsymbol{f}^{EL} - \boldsymbol{f}^{LfLE}$ is the E-L force density exerted by the electromagnetic wave on atoms of a host medium. It corresponds to the force $(\boldsymbol{P}\cdot\nabla)\boldsymbol{E} + \mu_0\frac{\partial \boldsymbol{P}}{\partial t}\times\boldsymbol{H}$ in the case of a dielectric host medium, which has been interpreted as the density of the Lorentz force exerted by the electromagnetic wave on dipoles [7], [17] of a host medium.

In accordance with interconnection of alternative equations of conservation of the momentum, the force density $\boldsymbol{f}^{EL}$ can be formulated [9] using components of Minkowski's momentum conservation equation

$$\boldsymbol{f}^{EL} = \boldsymbol{f}^{Lf} + \boldsymbol{f}^H + \frac{1}{2}\nabla(\boldsymbol{E}\cdot\boldsymbol{P} + \mu_0\boldsymbol{H}\cdot\boldsymbol{M}) + \boldsymbol{f}^A, \tag{14}$$

where $\delta_{ij}(\boldsymbol{E}\cdot\boldsymbol{P} + \mu_0\boldsymbol{H}\cdot\boldsymbol{M})$ is a part of the Minkowski stress tensor, which depends on the medium polarization and magnetization. The sum of components $\boldsymbol{f}^H + \frac{1}{2}\nabla\cdot\delta_{ij}(\boldsymbol{E}\cdot\boldsymbol{P} + \mu_0\boldsymbol{H}\cdot\boldsymbol{M})$ present in the formula (14) yields the expression $\frac{1}{2}\varepsilon_0(\varepsilon - 1)\nabla E^2 + \frac{1}{2}\mu_0(\mu - 1)\nabla H^2$, which does not depend on derivatives of constitutive parameters. It means that the E-L force does not depend on medium inhomogeneity associated with the radiation pressure. Using the formula $\frac{1}{2}\nabla a^2 = (\boldsymbol{a}\cdot\nabla)\boldsymbol{a} + \boldsymbol{a}\times(\nabla\times\boldsymbol{a})$ [23] and equations of the macroscopic electromagnetic field for transformation of the latter expression into the formula (13), components of the Abraham force density $\boldsymbol{P}\times\frac{\partial \boldsymbol{B}}{\partial t}$ and $\frac{1}{c^2}\frac{\partial \boldsymbol{E}}{\partial t}\times\boldsymbol{M}$ are eliminated using components $\boldsymbol{P}\times(\nabla\times\boldsymbol{E})$ and $\frac{\mu_0}{\varepsilon}\boldsymbol{M}\times(\nabla\times\boldsymbol{H})$ linked with the tensor $\delta_{ij}(\boldsymbol{E}\cdot\boldsymbol{P} + \mu_0\boldsymbol{H}\cdot\boldsymbol{M})$.

Physical significance of the force density $\mu_0\frac{\partial \boldsymbol{P}}{\partial t}\times\boldsymbol{H}$, (which represents the component of the E-L force density), is clear because it corresponds to the Lorentz type force associated with the displacement current depending on medium polarization. This type of interaction has been experimentally verified in experimental studies [14] - [15] of interaction of a dielectric body with quasi-static electric and magnetic fields. In contrast, the term $\boldsymbol{P}\times\frac{\partial \boldsymbol{B}}{\partial t}$ related with the Abraham force, which is present in the case of the equation (14), is absent in the equation (13). Nevertheless, the latter term is responsible for description of interaction of dipoles with the vortex electric field, which is caused by the change of the magnetic field in the case of the electromagnetic wave in a host medium in accordance with the Maxwell equations. On the other hand, such interaction is not significant in the case of interaction of the electromagnetic wave with electrically small dipoles, which do not constitute a continuous medium of a sufficient electrical size for exhibition of wave properties.

While E-L force density (13) does not include components of the Abraham force, which depend on time derivatives of field associated parameters $\frac{\partial \boldsymbol{B}}{\partial t}$ and $\frac{\partial \boldsymbol{E}}{\partial t}$, the total momentum is characterized in the context of interconnection of alternative approaches by elimination of components related with a part of the momentum described by the equation (11). Thus, discussed alternative approaches differently interpret form and nature of the electromagnetic wave momentum associated with the medium polarization and magnetization.

Analogously, it is possible to transform the combination of force densities on the right-hand side of the equation (14) into a force density, which depends on components $\boldsymbol{P} \times \frac{\partial \boldsymbol{B}}{\partial t}$ and $\frac{1}{c^2} \frac{\partial \boldsymbol{E}}{\partial t} \times \boldsymbol{M}$, while components $\mu_0 \frac{\partial \boldsymbol{P}}{\partial t} \times \boldsymbol{H} + \frac{1}{c^2} \boldsymbol{E} \times \frac{\partial \boldsymbol{M}}{\partial t}$ related with E-L force density are eliminated using the tensor $\delta_{ij}(\boldsymbol{E} \cdot \boldsymbol{P} + \mu_0 \boldsymbol{H} \cdot \boldsymbol{M})$ through the use of transformations, which are similar to ones discussed above.

Interconnection of alternative momentum conservation equations and their apparent functional equality can be explained by equivalence of transformations, which preserve the material derivative of the density of the full momentum of a closed medium-field system in agreement with the law of conservation of the momentum. While the electromagnetic momentum (or its part in the case of Minkowski's approach) is associated with Abraham's momentum in accord with the alternative approaches, adequate modification and interpretation of the equation (11) is a focus of interconnection. It is worth noting in this connection that the full momentum of a closed medium-field system is not changed while a surrogate equation $\frac{1}{2}(\boldsymbol{f}^A + \nabla \cdot \bar{\bar{T}}^A) = -\frac{1}{2}(\boldsymbol{f}^H - \boldsymbol{f}^{Af})$ is used instead of the equation (11) in the context of interconnection of alternative approaches. The above modification of the equation (11) is not reduced to redistribution of the electromagnetic and the medium components of momentum conservation equation. The time derivative of the density of electromagnetic wave momentum, $\frac{1}{2} \boldsymbol{f}^A$, related with the medium polarization and magnetization is decreased by two times in comparison with the equation (11), and it results in another definition of the ponderomotive force in accord with properties of the total momentum. Summing the surrogate equation, which is equivalent to the equation (11) in the context of interconnection, and equation (9) for the electromagnetic component of optical momentum, we obtain an equation of conservation of the momentum applicable in the case of the total momentum interpreted as the optical momentum. Accordingly, interconnection of the alternative approaches is not reduced to redistribution of medium and electromagnetic components of optical momentum in contradiction with assumed conditions of suggested justification [8] of functional equivalence of the alternative energy-momentum tensors of the electromagnetic field in a host medium.

### 3. RADIATION PRESSURE IN THE CONTEXT OF INTERCONNECTION

#### 3.1. FLUXES OF THE ENERGY AND MOMENTUM

While oscillations of the radiation pressure at optical frequencies are not detectable [11], interconnection of alternative approaches has been interpreted as functional equivalence of Abraham's and Minkowski's concepts of optical momenta in framework of the averaged radiation pressure. It has been assumed that the averaged radiation pressure is determined by the Helmholtz force in the case of rival Abraham's and Minkowski's approaches because the Abraham force linked with the time derivative of the Abraham momentum is vanishing upon time averaging.

In order to clarify interconnection of the alternative approaches, we examine the time averaged equation (11), $\langle \boldsymbol{f}^A + \nabla \cdot \bar{\bar{T}}^A \rangle = \langle \boldsymbol{f}^{Af} - \boldsymbol{f}^H \rangle$, which properties are essential for discussion of the component of optical momentum related with the medium polarization and magnetization. Taking into account a following equation $\boldsymbol{f}^A + \nabla \cdot \bar{\bar{T}}^A \big|_{n(r)=const} = 0$ obtained using the equation (11), one can derive a relation $\langle \nabla_{n(r)} \cdot \bar{\bar{T}}^A \rangle = \langle \boldsymbol{f}^{Af} - \boldsymbol{f}^H \rangle$, where the differential operator $\nabla_{n(r)}$ is applied to the coordinate dependence of the refractive index of a host medium. Examination of above equations demonstrates that contribution of the Abraham force, which is vanishing upon time averaging, has no impact on the (average) radiation pressure in contradiction with mentioned assumptions [9], [11].

It is worth noting that the material derivative of the full momentum of the electromagnetic wave interacting with a host medium is preserved in the case of transformations linking alternative equations of conservation of the momentum in general case. In this context debates of properties of the averaged Abraham force in separation from another part of the material derivative of optical momentum, $\nabla \cdot \bar{\bar{T}}^A$, which represents a volume force density, are meaningless.

Medium-field interaction in the case of an inhomogeneous medium is determined by the rate of change of the material derivative of optical momentum in agreement with Newton's laws and properties of translational invariance of the optical momentum in regard to a homogenous medium. At the same time, properties of ponderomotive forces related with transport of optical momentum components associated with medium-field interaction in the case of a homogeneous medium are less obvious in accordance with above discussion of properties of the Abraham force. Nevertheless, even though the time derivative of optical momentum is vanishing upon averaging, dependence of medium-field interaction on the flux of optical momentum is essential. To this end, the averaged radiation pressure on an interface between homogeneous dielectrics is determined below through the use of integral equations of conservation of fluxes of the energy and the momentum of the electromagnetic wave.

Using the energy current density of the electromagnetic field $S$, one can obtain an equation of conservation of the average energy current density for the incident wave in the case of normal incidence of light on an interface between non-dispersive dielectrics with negligible losses

$$\langle S \rangle = \frac{(n_1-n_0)^2}{(n_1+n_0)^2}\langle S \rangle + \frac{4n_1 n_0}{(n_1+n_0)^2}\langle S \rangle, \tag{15}$$

where, respectively, the first and second summands on the right-hand side of the equation are energy current densities of the reflected and transmitted waves. Partial reflection of the incident wave at the interface between dielectrics is taken into account making use of Fresnel's coefficients of reflection and transmission [23].

The average flux density of the linear momentum of an electromagnetic wave in free space is determined by $\frac{\langle S \rangle}{c}$ ratio. The average flux density of the linear momentum of an electromagnetic wave in host medium is described by an expression $x(n)\frac{\langle S \rangle}{c}$. The parameter $x(n)$ is ratio of momenta of the electromagnetic wave in the case of a host medium and free space, which depends on the refractive index of a host medium in accordance with alternative approaches. Making use of the energy flux parameters used in the equation (15), we obtain flux densities of optical momenta of the incident $x(n_0)\frac{\langle S \rangle}{c}$ and reflected $-\frac{(n_1-n_0)^2}{(n_1+n_0)^2}x(n_0)\frac{\langle S \rangle}{c}$ waves in a dielectric with the refractive index $n_0$ as well as the flux density of the optical momentum of the transmitted wave $\frac{4n_1 n_0}{(n_1+n_0)^2}x(n_1)\frac{\langle S \rangle}{c}$ in a dielectric with the refractive index $n_1$.

The material derivative of the optical momentum of the electromagnetic wave in a homogenous medium with negligible losses is equal zero in general case in accordance with the Section 2. Otherwise, the optical momentum of the electromagnetic wave would change in absence of dissipation or amplification in a host medium in contradiction with the law of conservation of the momentum. Discontinuity of the normal optical momentum flux of the incident wave caused by reflection and transmission of light at the interface between dielectrics is balanced by accompanying fluxes of mechanical momenta to provide conservation of the density of the full momentum of a closed medium-field system

$$x(n_0)\frac{\langle S\rangle}{c} = -\frac{(n_1-n_0)^2}{(n_1+n_0)^2}x(n_0)\frac{\langle S\rangle}{c} + 2\frac{(n_1-n_0)^2}{(n_1+n_0)^2}x(n_0)\frac{S}{c} + \frac{4n_1n_0}{(n_1+n_0)^2}x(n_1)\frac{\langle S\rangle}{c}$$
$$-\frac{4n_1n_0}{(n_1+n_0)^2}(x(n_1)-x(n_0))\frac{\langle S\rangle}{c}.\tag{16}$$

The first and third terms on the right-hand side of the equation, which correspond to flux densities of optical momenta of the reflected and transmitted waves, determine the second and fourth terms responsible for the radiation pressure. The second term represents the flux density of the mechanical momentum caused by reflection. The fourth term is the flux density of the mechanical momentum balancing the change of the optical momentum of the transmitted wave passing through the interface. The radiation pressure corresponds to the flux density of the mechanical momentum associated with the second and fourth terms.

Because mechanical waves associated with the radiation pressure, are comparatively slow, they do not provide essential contribution to the average energy current. Taking sums of the first and second as well as third and fourth terms, the balance equation (16) for fluxes of optical and mechanical momenta is reduced to the equation (15). Accordingly, the equation (16) associated with the alternative concepts of optical momenta is consistent with the equation (15) of conservation of the average energy current density of the electromagnetic wave.

It is instructive to make sure that the integral equation (16) complies with differential equations of conservation of the momentum. Moving the first three terms located on the right-hand side of the equation (16) to the left-hand side, one can obtain an equation for the transmitted wave

$$\frac{4n_1n_0}{(n_0+n_1)^2}\frac{\langle S\rangle}{c}x(n_1) - \frac{4n_1n_0}{(n_0+n_1)^2}\frac{\langle S\rangle}{c}x(n_0) = -\frac{4n_1n_0}{(n_0+n_1)^2}\frac{\langle S\rangle}{c}(x(n_0)-x(n_1)).\tag{17}$$

In the case of an infinitesimal layer of an inhomogeneous medium associated with an interface between dielectrics with close refractive indices $n_0$ to $n_1$, the refractive index is slowly changed from $n_0$ to $n_1$. Because $\lim_{n_1\to n_0}\frac{4n_1n_0}{(n_0+n_1)^2}=1$ in this case, the energy current density of the incident wave in a dielectric with the refractive index $n_0$ is transformed into the energy current density of the transmitted wave in a dielectric with the refractive index $n_1$.

The change of the average flux density of the optical momentum $(n_1-n_0)\frac{\langle S\rangle}{c}$ of the wave passing through the interface layer corresponds to the change of the radiation pressure $(n_0-n_1)\frac{\langle S\rangle}{c}$ in accordance with the equation (17) in the case $x(n)=n$. While the change of the radiation pressure represents an integral parameter associated with the layer of inhomogeneous medium, an average force density $-\frac{\langle S\rangle}{c}\nabla n$ is a corresponding local parameter. Using the energy current density in a dielectric in the form $S=\varepsilon_0 ncE^2$, one can demonstrate that the expression $-\frac{S}{c}\nabla n$ represents the Helmholtz force density in the case of the electromagnetic wave, which can be transformed into the formula $-\frac{\varepsilon_0}{2}E^2\nabla n^2$ associated with the electric field in accordance with the formula (5) for $\boldsymbol{f}^H$.

The change of the averaged flux density of the optical momentum on the left-hand side of the equation (17) is transformed into the expression $\langle\frac{\partial}{\partial t}\boldsymbol{D}\times\boldsymbol{B}+\nabla\frac{n}{c}S\big|_{n=const}\rangle$, where it is taken into account that $\frac{\partial}{\partial t}\boldsymbol{D}\times\boldsymbol{B}+\nabla\frac{n}{c}S\big|_{n=const}=0$ because functions $\frac{\partial}{\partial t}\boldsymbol{D}\times\boldsymbol{B}$ and $\nabla\frac{n}{c}S\big|_{n=const}$ oscillate with the antiphase shift in the case of a medium with negligible losses. Accordingly, the integral of a time-averaged differential equation (5) across the layer of inhomogeneous medium with negligible losses $\int \langle \frac{\partial}{\partial t} \boldsymbol{D} \times \boldsymbol{B} + \nabla \cdot \bar{\bar{T}}^{Min} \rangle d\boldsymbol{r} = -\int \langle \boldsymbol{f}^H \rangle d\boldsymbol{r}$ corresponds to the integral equation (17) in the case $x(n) = n$.

### 3.2. RADIATION PRESSURE

Using the equation (16) in the case of the optical momentum having form of Minkowski's momentum, the average pressure on the interface between dielectrics is determined by a following formula

$$\langle p^M \rangle = \frac{2(n_0 - n_1)}{(n_0 + n_1)} n_0 \frac{\langle S \rangle}{c}. \tag{18}$$

Making use of the formula (18) in the case of the infinitesimal layer of an inhomogeneous medium of the width $\Delta x$, one can connect the change of the average radiation pressure $\langle \Delta p^M \rangle = -\Delta n \frac{\langle S \rangle}{c}$ with the change of the refractive index $\Delta n = n_1 - n_0$. Calculating the rate of change of the average radiation pressure, the force density $\frac{\langle \Delta p^M \rangle}{\Delta x} = -\frac{\langle S \rangle}{c} \nabla_x n$, which corresponds to the averaged Helmholtz force density, is obtained.

The averaged radiation pressure (18) on the interface between dielectrics is determined by the sum of averaged fluxes of mechanical momenta caused by transmission and reflection of the incident wave in accordance with discussion of the equation (16). On the other hand, the force density associated with the radiation pressure $\boldsymbol{f}^H = -\frac{S}{c} \nabla n$ is obtained making use of the spatial rate of change of the flux density of Minkowski's momentum related with medium inhomogeneity in accordance with the previous subsection. It demonstrates that the radiation pressure does not directly depend on the Abraham force, which determines the time derivative of the medium related momentum accompanying the electromagnetic wave in a host medium.

The averaged radiation pressure on the interface between non-dispersive dielectrics caused by the flux density of Abraham's momentum is determined using the equation (16) in the case $x(n) = \frac{1}{n}$ while the photon momentum in a host medium is presented in the form $\frac{\hbar \omega}{cn}$ in this case

$$\langle p^{Af} \rangle = -\frac{2(n_0 - n_1)}{(n_0 + n_1)} \frac{1}{n_0} \frac{\langle S \rangle}{c}. \tag{19}$$

Making use of the formula (19) in the case of the infinitesimal difference of the refractive index $\Delta n = n_1 - n_0$ associated with medium layer of the width $\Delta x$, one can obtain $\langle \Delta p^{Af} \rangle = \frac{\Delta n}{n^2} \frac{\langle S \rangle}{c}$. Using the rate of change of the averaged radiation pressure $\frac{\langle \Delta p^{Af} \rangle}{\Delta x}$, corresponding force density is presented in a following form

$$\langle \boldsymbol{f}^{Af} \rangle = -\frac{\langle \boldsymbol{f}^H \rangle}{n^2}. \tag{20}$$

Besides, the force density $\boldsymbol{f}^{Af}$ is obtained in a form $-S\nabla \frac{1}{cn}$ making use of the spatial rate of change of the flux density of the Abraham momentum in an inhomogeneous host medium.

Using equation (16) in the case $x(n) = \frac{1}{2}\left(n - \frac{1}{n}\right)$, when optical momentum is interpreted as the total momentum, the radiation pressure on the interface between non-dispersive dielectrics is described by a formula

$$\langle p^t \rangle = \frac{2(n_0-n_1)}{(n_0+n_1)} \frac{1}{2} \left(n_0 - \frac{1}{n_0}\right) \frac{\langle S \rangle}{c}. \tag{21}$$

Making use of the formula (21) in the case of the infinitesimal difference of the refractive index $\Delta n = n_1 - n_0$ associated with medium layer of the width $\Delta x$, we obtain $\langle \Delta p^t \rangle = -\frac{\Delta n}{2}\left(1 - \frac{1}{n^2}\right)\frac{\langle S \rangle}{c}$. Using the latter formula, one can obtain a force density $\langle f^t \rangle = \frac{\langle \Delta p^t \rangle}{\Delta x}$ applicable in the case of the total momentum associated with the electromagnetic wave in a host medium

$$\langle f^t \rangle = \frac{\langle f^H \rangle}{2}\left(1 - \frac{1}{n^2}\right). \tag{22}$$

It is worth noting that, formula (22) is also obtained using results for the radiation pressure associated with Abraham's and Minkowski's momenta making use of interconnection of the alternative approaches. While the total momentum is presented by the half of the sum of Abraham's and Minkowski's momenta, the force density $f^t$ satisfies a relation $f^t = \frac{1}{2}(f^H + f^{Af})$, which replicates the relation linking the total, Minkowski's, and Abraham's momenta.

Because the E-L force density (13) does not depend on medium inhomogeneity in accordance with examination of derivation procedure meaning in subsection 2.3, E-L force density is not adequate in the case of interpretation of the total momentum as optical momentum of the electromagnetic wave in a host medium. On the other hand, it is worth noting that E-L force can be easily corrected in this context by means of inclusion of the force density (22) related with medium inhomogeneity using equivalent transformation of E-L equation of conservation of the momentum.

It has been demonstrated above that vanishing of the average Abraham force and the average time derivative of the Abraham momentum do not mean that the radiation pressure does not depend on fluxes of these momenta, which accompany the electromagnetic wave in a host medium. Accordingly, contribution to the radiation pressure related with a component of the averaged flux density of optical momentum corresponding to the Abraham force is calculated using formulas (18) and (19) obtained making use of Minkowski's and Abraham's momenta

$$\langle p^{af} \rangle = \frac{2(n_0-n_1)}{(n_0+n_1)}\left(n_0 + \frac{1}{n_0}\right)\frac{\langle S \rangle}{c}. \tag{23}$$

Modifying formula (23) for the case of infinitesimal layer of an inhomogeneous medium, one can obtain that $\langle \Delta p^{af} \rangle = -\Delta n\left(1 + \frac{1}{n^2}\right)\frac{\langle S \rangle}{c}$. Using the latter formula, we obtain the force density $\langle f^{af} \rangle = \frac{\langle \Delta p^{af} \rangle}{\Delta x}$ in a form,

$$\langle f^{af} \rangle = \langle f^H \rangle \left(1 + \frac{1}{n^2}\right), \tag{24}$$

where $f^{af} = f^H - f^{Af}$ is the force density exerted in the case of an inhomogeneous medium due to change of the flux of the component of optical momentum, which time derivative is determined by the Abraham force.

Thus, in compliance with formulas (18) - (24), the radiation pressure is determined by forces related with medium inhomogeneity, which forms are linked with dependences of alternative optical momenta on the refractive index of a host medium, whereas neither the Abraham force nor its components

## 4. INTERCONNECTION IN THE LIGHT OF MICROSCOPIC DESCRIPTION

### 4.1. SIGNIFICANCE OF MICROSCOPIC DESCRIPTION

Interconnection of alternative approaches to definition of optical momentum based on equations of macroscopic electromagnetic field has been studied using microscopic models of medium-field interaction for a long time. About fifty years ago, interconnection of macroscopic approaches was interpreted by Ginsburg [5] using an analogy between inertial properties of electromagnetic and mechanical waves suggested by Keldysh. In accordance with that, propagation of electromagnetic and sound waves is not accompanied by displacement of mass of a host medium. Accordingly, the momentum of sound waves is equal to zero if the relativistic mass $\frac{\hbar\omega}{c^2}$ corresponding to the energy of the phonon $\hbar\omega$, which represents the quantum of the sound wave, is not taken into account. It is obtained upon quantizing that the phonon has the energy $\hbar\omega$ and zero momentum while the momentum $\frac{\hbar\omega}{c^2}\boldsymbol{s}$, where $\boldsymbol{s}$ is the velocity of sound here, associated with the energy is neglected. The statement that the phonon has momentum (for example, in the case of its emission by the electron) equal to $\hbar\boldsymbol{k} = \frac{\hbar\omega}{s}\frac{\boldsymbol{k}}{k}$, where $\boldsymbol{k}$ is the wave vector, means that the medium lattice obtains this momentum. Nothing is changed in application of the laws of conservation of the momentum and the energy, if this momentum is attributed to the phonon. The same applies in the case of the electromagnetic wave in a host medium, while Abraham's momentum $\frac{\hbar\omega}{c^2}\boldsymbol{v}_s$ of the photon has similar meaning as the "true" momentum $\frac{\hbar\omega}{c^2}\boldsymbol{s}$ of the phonon.

In accordance with the discussed hypothesis, the mechanical momentum, which time derivative is determined by the Abraham force, does not accompany Abraham's momentum, and motion of a host medium is determined by equations of elasticity and hydrodynamics. Accordingly, Abraham's momentum $\frac{\hbar\omega}{c^2}\boldsymbol{v}_s$ is the optical momentum, whereas conservation of the momentum during emission and absorption of photons is described using Minkowski's momentum. This explanation justifies significance of both Abraham's and Minkowski's momenta of photons on basis of presumed mechanical properties of the momentum related with the Abraham force. Such interpretation is based on an assumption that the momentum, the time derivative of which is determined by the Abraham force, is left in a host medium as a result of medium-field interaction. However, this is not consistent with the fact that the time averaged Abraham force is vanishing.

A concept of the "crystal momentum" or the pseudomomentum $\hbar\boldsymbol{k}$ of the electromagnetic wave in a host medium associated with elementary excitations $\hbar\omega$ has been explored by Gordon [7] later. It was suggested that Minkowski's form of the momentum corresponds to the "crystal momentum" associated with de Broglie's principle because the ratio of the energy of the electromagnetic field to Minkowski's momentum corresponds to the phase velocity. The "crystal momentum" of the electromagnetic wave in a host medium has been linked with Abraham's momentum and medium-field interaction in contrast with Ginsburg's interpretation. However, an assumed model of medium-field interaction associated with the total momentum in the context of alternative optical momenta is not consistent as with the mentioned above form of the "crystal momentum".

The problem of interconnected momenta of the electromagnetic wave interacting with a host medium reappeared in discussion of the Abraham-Minkowski controversy a quarter of a century later when properties of translational invariance of the electromagnetic wave in a host medium were discussed [12] in the context of Noether's theorem. Using the Amperian formulation [10] of equations of the macroscopic electromagnetic field in the case of a dielectric medium, the later interpretation of medium-field interaction was also linked with the Lorentz type force in accord with Balazs' interpretation [4] of inertial properties of photons in a host medium.

**4.2. BASICS OF MICROSCOPIC DESCRIPTION**

While alternative concepts of the optical momentum are associated with different definitions of ponderomotive forces, the Lorentz force, which significance in the case of microscopic description of medium-field interaction is well-known, has been used for justification of some macroscopic approaches. E.g., E-L force density in the case of a dielectric medium, $(\boldsymbol{P} \cdot \nabla)\boldsymbol{E} + \mu_0 \frac{\partial \boldsymbol{P}}{\partial t} \times \boldsymbol{H}$, is interpreted as the time derivative of the density of the mechanical momentum associated with the electromagnetic wave [14], [15] using similarity of forms of the E-L and Lorentz forces. However, the Lorentz force density, $(\boldsymbol{p} \cdot \nabla)\boldsymbol{E} + \frac{\partial \boldsymbol{p}}{\partial t} \times \boldsymbol{B}$ [7], [17] - [19], just describes the rate of change of the mechanical momentum of an electrically small dipole with the dipole moment $\boldsymbol{p}$ in the electromagnetic field.

Interaction of the electromagnetic wave with a host medium is not reduced to action of the electromagnetic field on an electrically small body while notion of optical momentum in a host medium is not applicable in latter conditions. Besides, $(\boldsymbol{P} \cdot \nabla)\boldsymbol{E}$ component of E-L's and Lorentz's forces is zero in the case of the plain electromagnetic wave in a homogeneous medium. On the other hand, the Abraham force density, $\frac{\partial}{\partial t}\{\boldsymbol{D} \times \boldsymbol{B} - \varepsilon_0\mu_0 \boldsymbol{E} \times \boldsymbol{H}\}$, which is presented in the case of a dielectric medium using equations of the macroscopic electromagnetic field in a form $\mu_0 \frac{\partial \boldsymbol{P}}{\partial t} \times \boldsymbol{H} + (\nabla \times \boldsymbol{E}) \times \boldsymbol{P}$, depends on the displacement current as well as coordinate derivatives of the vortex electric field.

Properties of medium-field interaction are clarified taking into account that the medium polarization and magnetization depend on "mechanical" oscillatory motion of atoms' electrons induced by the electromagnetic field, consideration of which is essential for description of electromagnetic properties of a medium in accord with the Lorentz theory of dispersion. Besides, it is necessary to take into account re-radiation of the electromagnetic waves by atoms' electrons, which interference with the primary wave explains phase, group, and energy velocities of light in a medium. In contrast with macroscopic approaches associated with "mechanical" components of the optical momentum, re-radiation of electromagnetic waves means that the energy of the electromagnetic wave is transferred to atoms' electrons, which emit secondary waves. Transfer of the field energy is accompanied by transfer of "mechanical" and "electromagnetic" components of optical momentum, which are linked with "mechanical" motion of atoms' electrons and secondary radiation, respectively. In this connection, $\mu_0 \frac{\partial \boldsymbol{P}}{\partial t} \times \boldsymbol{H}$ and $(\nabla \times \boldsymbol{E}) \times \boldsymbol{P}$ are components of the Abraham force density associated with transfer of "mechanical" and "electromagnetic" components of atoms' momenta. Components of the Abraham force density, $\frac{1}{c^2}\boldsymbol{E} \times \frac{\partial \boldsymbol{M}}{\partial t} + \frac{1}{c^2}\frac{\partial \boldsymbol{E}}{\partial t} \times \boldsymbol{M}$, related with the medium magnetization are interpreted as time derivatives of the densities of the "mechanical", $\frac{1}{c^2}\boldsymbol{E} \times \frac{\partial \boldsymbol{M}}{\partial t}$, and the "electromagnetic", $\frac{1}{c^2}\frac{\partial \boldsymbol{E}}{\partial t} \times \boldsymbol{M}$, momenta accompanying the electromagnetic wave similar to the case associated with the electromagnetic wave in a dielectric.

Furthermore, propagation of the electromagnetic wave in a host medium has to be taken into account using the Lagrangian derivative of the medium component of optical momentum in accordance with discussion of equation (11). The Abraham and E-L forces determine time derivatives of densities of rival momenta of the electromagnetic wave related with the medium polarization and magnetization, whereas a generalized Lorentz force associated with the Abraham force is determined by the expression $\boldsymbol{f}^A + \nabla \cdot \bar{\bar{T}}^A$ in accord with equation (11). This form corresponds to the ponderomotive force, which takes into account as reversible transfer "mechanical" and "electromagnetic" components of the optical momentum between atoms of a host medium as well as their transport in a host medium.

Presence of mechanical components of optical momentum contradicts to discussed microscopic description, in accordance to which oscillating atoms' electrons obtain and radiate the electromagnetic energy. The "mechanical" momentum of bound electrons and corresponding "electromagnetic" momentum associated with the polarization and the magnetization of a host medium in the field of the electromagnetic wave are described by Maxwell equations rather than equations of hydrodynamics and elasticity applicable in the case of mechanical motion of atoms of a host medium. In agreement with discussion of interconnection of alternative macroscopic approaches in Section 2, the time derivative of the sum of the densities of the "mechanical" and the "electromagnetic" momenta is determined by the density of the Abraham force while magnitudes of time derivatives of the densities of the "mechanical", $\frac{\partial \boldsymbol{P}}{\partial t} \times \boldsymbol{B} + \frac{1}{c^2} \boldsymbol{E} \times \frac{\partial \boldsymbol{M}}{\partial t}$, and the "electromagnetic", $\boldsymbol{P} \times \frac{\partial \boldsymbol{B}}{\partial t} + \frac{1}{c^2} \frac{\partial \boldsymbol{E}}{\partial t} \times \boldsymbol{M}$, components of momenta [22] are equal in the case of a host medium with negligible losses.

On the other hand, E-L's force is adequate in the case of quasi-static fields because in such case the momentum is transferred to an electrically small body. Accordingly, this momentum is mechanical in contrast with the case of "mechanical" and "electromagnetic" components of optical momentum of the electromagnetic wave in a host medium. Medium-field interaction associated with alternative optical momenta is reduced to E-L's force while optical momentum concept is not applicable in the case of quasi-static fields. In this context, taxonomy of combinations of ponderomotive forces and associated optical momenta discussed in Sections 2 and 3 assumes significance of the generalized Lorentz force in the case of the electromagnetic wave in a host medium.

In addition, it is necessary to emphasize significance of microscopic description of propagation of light between atoms of a host medium at the velocity of light in empty space in accordance with Einstein's postulate. In this case, propagation of the electromagnetic wave energy is explained [24] - [26] in agreement with relativistic causality by superposition of the primary wave with secondary electromagnetic waves reradiated by a host medium.

### 4.3. THE "EINSTEIN BOX" THOUGHT EXPERIMENT

Insufficiency of macroscopic description of optical momentum manifests itself by alternative equations of conservation of the momentum associated with the Abraham-Minkowski controversy. It explains interest to so-called "Einstein box" thought experiment [6], [17], in the case of which optical momentum of the electromagnetic wave in a host medium has been defined using assumptions in regard to inertial properties of photons in a host medium.

Making use of the relativistic mass associated with the energy of the electromagnetic field for description of inertial properties of the plane electromagnetic wave going through a transparent slab, it has been suggested by Balazs [4] that Abraham's momentum represents the optical momentum of the electromagnetic wave in a host medium. In accord with definition of the kinetic momentum of a body [16], the linear momentum of a photon of the electromagnetic wave is determined by the formula

$$\boldsymbol{P} = m\boldsymbol{v}_s, \quad (25)$$

where $m = \frac{\hbar \omega}{c^2}$ is the relativistic mass of a photon, and $\boldsymbol{v}_s$ is the energy velocity, which is equal to the group velocity [12] in the case of a medium with negligible losses as well as to the phase velocity $\boldsymbol{v}_{ph}$ in the case of a non-dispersive medium without losses. If photon's energy $\hbar \omega$ is attributed to the electromagnetic momentum linked with the mass $m$, the momentum of a photon is presented in agreement with the formula (25) in a form $\frac{\hbar\omega}{cn_s}$, where $n_s = \frac{c}{v_s}$. The ratio of the energy of the electromagnetic wave to the magnitude of the optical momentum is determined in this case by the expression $\frac{mc^2}{mv_s}$.

Alternative macroscopic descriptions are consistent with the law of conservation of momentum in accordance with Section 3, whereas Balazs's interpretation of inertial properties of the electromagnetic wave in a host medium is not sufficient in the context of the Abraham-Minkowski controversy. The alternative ratio of the energy $\hbar\omega$ to optical momentum, which is equal to the phase velocity and consistent with de Broglie's principle in the case of the electromagnetic wave in a host medium, is obtained using Minkowski's form of photon's momentum, $\frac{n\hbar\omega}{c}$. Presenting Minkowski's form of photon's momentum in a form $m^* v_g$, inertial properties of photons are characterised [6] using Veselago's mass parameter $m^* = \frac{\hbar\omega}{v_{ph} v_g}$, which defines inertial properties taking into account the difference of phase and group velocities in contrast with [30]. In contradiction with Balazs' interpretation of inertial properties of the electromagnetic wave in a host medium, $m^*$ depends on medium properties.

It is worth noting that discussed interpretations of inertial properties, which are consistent with Einstein's relation for the mass-energy and de Broglie's principle, correspond to rival approaches. Rivalry of descriptions of inertial properties implies contraposition of Einstein's relation for the mass-energy and de Broglie's principle in the context of the Abraham-Minkowski controversy. It demonstrates deficiency of alternative macroscopic approaches as well as discussed interpretations of inertial properties of photons in a host medium.

The photon energy, $\hbar\omega$, is fully assigned to Abraham's momentum in accordance with Balazs' assumptions. As a result, medium related momentum transferred due to interaction with the electromagnetic wave is associated with mechanical motion of atoms, whose energy, linked with such kind of motion, is negligible [10]. However, combining of electromagnetic and mechanical components of optical momentum is inconsistent because propagation of electromagnetic waves in a medium is described by the Maxwell equations of the macroscopic electromagnetic field. Accordingly, Abraham's momentum interpreted as the optical momentum is a more reasonable option than the optical momentum comprised of the Abraham momentum and a mechanical momentum, (the time derivative of which is determined by either the Abraham or a Lorentz-like force). On the other hand, even though concepts associated with mechanical components of optical momentum are deficient, they take into account electromagnetic wave's interaction with a host medium in contrast with Abraham's and Minkowski's concepts of electromagnetic momenta.

A more rational interpretation of Abraham's form $\frac{\hbar\omega}{cn_s}$ of the electromagnetic momentum of the photon follows from microscopic description of light propagation. While the velocity of light in vacuum is used for description of transport of the electromagnetic momentum in accordance with Einstein's postulate [22], associated energy $\frac{\hbar\omega}{n_s}$ and relativistic mass $\frac{\hbar\omega}{c^2 n_s}$ attributed to Abraham's kinetic momentum of the photon in a medium is decreased in $n_s$ times in comparison with values related with Balazs' interpretation. Accordingly, the energy per a photon reversibly transferred to atom's electrons is determined by the expression $\hbar\omega - \frac{\hbar\omega}{n_s}$. As a result of energy exchange between the electromagnetic wave and a host medium, propagation of the electromagnetic wave in a medium can be explained by superposition of the primary wave with secondary electromagnetic waves reradiated by atoms of a host medium in accordance with microscopic description.

Because "electromagnetic" and "mechanical" components of the momenta related with the medium polarization and magnetization are interconnected in accordance with microscopic description, a special significance of Minkowski's form of the momentum, $n\frac{\hbar\omega}{c}$, inevitably follows. In this context, the photon momentum, the time derivative of which is defined by the Abraham force, is determined by expression $n\frac{\hbar\omega}{c} - \frac{\hbar\omega}{cn_s}$ using Abraham's and Minkowski's forms of photon's momentum in a host medium.

### 4.4. INERTIAL PROPERTIES AND THE PSEUDOMOMENTUM

To clarify inertial properties of the electromagnetic wave in a host medium in the context of the interconnection, we use an equation of conservation of the momentum, which describes medium-field interaction in a closed medium-field system in framework of alternative equations,

$$\frac{d\boldsymbol{p}}{dt} = -\boldsymbol{f}^* - \boldsymbol{f}^{loss} - \boldsymbol{f}^{sf}, \qquad (26)$$

where $\boldsymbol{p}$ is the density of the field momentum, $\boldsymbol{f}^*$ is a force density, which determines the rate of reversible transfer of the momentum between the electromagnetic wave and a host medium, $\boldsymbol{f}^{loss}$ is the volume rate of irreversible losses associated with dissipation of the electromagnetic wave momentum in a host medium, and $\boldsymbol{f}^{sf}$ is the ponderomotive force density related with medium inhomogeneity and radiation pressure.

Integrating an equation $\frac{d\boldsymbol{p}}{dt} = -\boldsymbol{f}^*$, which represents equation (26) in the case of a homogeneous medium with negligible losses, one can introduce the density of optical pseudomomentum $\widetilde{\boldsymbol{p}}$ of the electromagnetic wave in a host medium

$$\widetilde{\boldsymbol{p}} = \boldsymbol{p} + \int \boldsymbol{f}^* dt. \qquad (27)$$

It corresponds to the sum of the density of the electromagnetic momentum $\boldsymbol{p}$ of electric and magnetic fields in a host medium and pseudomomentum density $\boldsymbol{p}^* = \int \boldsymbol{f}^* dt$, which is related with interaction of the electromagnetic wave with atoms of the medium.

While inertial properties of the electromagnetic wave in a host medium are described using the relativistic mass in accordance with Balazs's interpretation, pseudomomentum caused by medium-field interaction is ignored. Nevertheless, significance of interaction of the electromagnetic field with a medium is obvious due to dependence of the field equations on the medium polarization and magnetization, so that $\boldsymbol{p}^* \neq 0$ in general case, even though Abraham's and Lorentz's forces are vanishing upon averaging. At the same time, recognition of the fact that $\boldsymbol{p}^* \neq 0$ is not sufficient for unique definition of pseudomomentum because equations (26) and (27) are consistent with alternative interpretations of medium-field interaction.

In the case of a homogeneous medium with negligible losses, the equation of conservation of momentum (26) is reduced to equation $\frac{d\widetilde{\boldsymbol{p}}}{dt} = 0$, which describes conservation of optical pseudomomentum. Because $\frac{d\widetilde{\boldsymbol{p}}}{dt} - \frac{d\boldsymbol{p}}{dt} = 0$ and $\frac{d\widetilde{\boldsymbol{p}}}{dt} - \frac{d\boldsymbol{p}^*}{dt} = 0$, equation $\frac{d\widetilde{\boldsymbol{p}}}{dt} = 0$ can be substituted by equations of conservation of comprising components of optical pseudomomentum, $\frac{d\boldsymbol{p}}{dt} = 0$ and $\frac{d\boldsymbol{p}^*}{dt} = 0$, in accordance with discussion of interconnection of alternative macroscopic approaches in the Section 2. In this context, different combinations of the electromagnetic momentum and pseudomomentum can be interpreted as the optical momentum of the electromagnetic wave in a host medium in accordance with alternative macroscopic approaches.

Equation of conservation of pseudomomentum, $\frac{d\boldsymbol{p}^*}{dt} = 0$, can be similarly interpreted in accord with interconnection of alternative macroscopic approaches. In this context, the latter equation is presented in a form $\frac{d\boldsymbol{p}^{*e}}{dt} = -\frac{d\boldsymbol{p}^{*m}}{dt}$ taking into consideration "mechanical" $\boldsymbol{p}^{*m}$ and "electromagnetic" $\boldsymbol{p}^{*e}$ components of pseudomomentum, $\boldsymbol{p}^* = \boldsymbol{p}^{*m} + \boldsymbol{p}^{*e}$. Furthermore, "mechanical" and "electromagnetic" components of pseudomomentum can be also separated. In accordance with that, a pair of equations $\frac{d\boldsymbol{p}^{*e}}{dt} = 0$ and $\frac{d\boldsymbol{p}^{*m}}{dt} = 0$ can be used instead of equation $\frac{d\boldsymbol{p}^*}{dt} = 0$. In this context, components related with the "electromagnetic" momentum can be eliminated in accord with description of interconnection of the total momentum with Minkowski's momentum in Section 2 because an equation $\boldsymbol{p}^{*e} = 0$ represents a particular solution of the equation $\frac{d\boldsymbol{p}^{*e}}{dt} = 0$.

In contrast, in accordance with microscopic description of the electromagnetic wave in a host medium, oscillations of atoms' electrons are linked with the secondary radiation, whose time lag in relation to the primary wave is responsible for dependence of propagation velocity of the electromagnetic wave on the refractive index. Besides, "mechanical", $\boldsymbol{p}^{*m}$, and "electromagnetic", $\boldsymbol{p}^{*e}$, components of the pseudomomentum are inherently interconnected. Taking into account these aspects for interpretation of the alternative macroscopic approaches, the pseudomomentum density is determined by the expression $(\boldsymbol{D} \times \boldsymbol{B} - \varepsilon_0\mu_0 \boldsymbol{E} \times \boldsymbol{H})$ while photon's pseudomomentum is determined by expression $\frac{\hbar\omega}{c}\left(n - \frac{1}{n_s}\right)$ using Abraham's and Minkowski's forms of photons' momenta. In this context, reversible transfer of "mechanical" and "electromagnetic" components of the pseudomomentum between atoms of a host medium is associated with the generalized Lorentz force discussed in subsection 4.1. It is worth noting that this force density is defined by means of the material derivative, $\boldsymbol{f}^* = \frac{d\boldsymbol{p}^*}{dt}$, rather than using the time derivative. Besides, conservation of the pseudomomentum $\boldsymbol{p}^* = \int \boldsymbol{f}^* dt$ of the electromagnetic wave in a homogeneous medium with negligible losses assumes that transfer of the pseudomomentum to atoms of host medium by means of the force density $\boldsymbol{f}^* = \frac{d}{dt}\left\{\boldsymbol{P} \times \boldsymbol{B} - \frac{1}{c^2}\boldsymbol{M} \times \boldsymbol{E}\right\}$ is reversible. Indeed, because the average of the Abraham force vanishes in the case of a homogeneous medium with negligible losses, the impulse left by the electromagnetic wave due to medium-field interaction is equal to zero in accordance with translational invariance of the electromagnetic wave in relation to a homogeneous medium without losses.

Using photon's electromagnetic momentum (25), the optical pseudomomentum $\widetilde{\boldsymbol{P}}$, which depends on medium-field interaction, is defined by a following formula

$$\widetilde{\boldsymbol{P}} = \widetilde{m}\boldsymbol{v}_s, \qquad (28)$$

where $\widetilde{m} = \frac{\hbar\omega}{vv_s}$ is the effective inertial mass of a photon, which includes the relativistic mass and depends on the change of inertial properties associated with reversible momentum transfer in process propagation of the electromagnetic wave in a host medium [22]. It is worth noting that this definition of the optical pseudomomentum is consistent with de Broglie's principle as well as description of inertial properties associated with Einstein's formula for the mass-energy.

The optical pseudomomentum was previously interpreted as the Minkowski momentum, the wave momentum, and the canonical momentum [8] - [12], [31]. However, the pseudomomentum depends on motion of atom electrons associated with medium-field interaction related with transfer of the electromagnetic energy to atoms, in contrast with the mechanical momentum of atoms, which depends on atoms' relativistic mass, and the electromagnetic momentum determined by the relativistic mass associated with the electromagnetic energy. For that reason, the difference between optical momentum and optical pseudomomentum is substantial, even though electromagnetic momenta associated with the electromagnetic field energy are comprising parts of discussed momenta and pseudomomenta. Because of this difference, combination of the electromagnetic momentum and the pseudomomentum of the electromagnetic wave in a host medium represents the optical pseudomomentum.

Inertial properties of the electromagnetic wave in a host medium are mistakenly interpreted in the case when the optical pseudomomentum is described as Minkowski's momentum. In such case, the change of the effective inertial mass $\widetilde{m} - m = \frac{\hbar\omega}{c^2}(nn_s - 1)$ of a photon in a host medium is implicitly interpreted as the relativistic mass, which accompanies the electromagnetic wave in accord with basic properties of the momentum. Using an "energy" parameter $\hbar\omega\, nn_s$, assumed significance of which follows from above interpretation of the mass parameter $\widetilde{m} = \frac{\hbar\omega}{c^2} nn_s$ as the "relativistic mass", to calculate the ratio of corresponding "relativistic energy" $\hbar\omega\, nn_s$ to Minkowski's momentum $\frac{\hbar\omega n}{c}$, the ratio $\frac{c^2}{v_s}$ is obtained, in accordance to which the pseudomomentum related with medium-field interaction is mistakenly categorised as the momentum associated with mechanical motion of atoms.

However, it is not correct to interpret the effective inertial mass related with the electromagnetic wave in a host medium as the relativistic mass while the latter mass is positively defined in general case while $\widetilde{m} - m$ parameter is negative if $n < \frac{1}{n_s}$. The effective inertial mass cannot be interpreted as the relativistic mass because such interpretation of the effective inertial mass is clearly incorrect in the case of negative refraction media. Dependence of the effective mass on the refractive index assumes significance of interaction of the electromagnetic wave with a host medium associated with transfer of the electromagnetic energy to atoms of a host medium. "Electromagnetic" and "mechanical" components of the optical pseudomomentum linked with the medium polarization and magnetization as well as the effective inertial mass are related with electron motion within atom shells rather than transport of medium substance. Accordingly, the energy of the electromagnetic wave in a medium is the energy of the corresponding electromagnetic field [24] - [26], which does not include the mass-energy of medium substance in accord with the concept of the optical pseudomomentum.

## 5. Conclusion

Interconnection of alternative equations of momentum conservation is discussed taking into consideration that possible modifications of the equations are not reduced to redistribution of medium and electromagnetic components of momenta of a closed medium-field system in contrast with previously debated assumptions. Potential functional equivalence of interconnected macroscopic approaches is examined in the case of description of the radiation pressure on an interface between dielectrics. It is demonstrated that averaging of equations of momentum conservation in the experimental context does not eliminate differences of alternative approaches. In accordance with obtained results, the average radiation pressure on an interface between dielectrics is determined by "electromagnetic" and "mechanical" components of alternative optical momenta described by their dependences on the refractive index of a host medium. In contrast with alternative macroscopic approaches, "electromagnetic" and "mechanical" components of optical momentum are interconnected because "mechanical" motion of bound electrons is responsible for secondary radiation, as it is demonstrated using microscopic description. In this context, Minkowski's form of the optical momentum is interpreted as the optical pseudomomentum, which inertial properties depend on the relativistic mass corresponding to the energy of the electromagnetic field as well as interaction of photons with a host medium. The material derivative of "electromagnetic" and "mechanical" components of optical momentum associated with medium-field interaction is described by the generalized Lorentz force, which is responsible for momentum exchange between the electromagnetic wave and a host medium. Thus, interconnection of alternative macroscopic approaches is clarified making use of microscopic description associated with consistent interpretation of interaction of the electromagnetic wave with a host medium.